\documentclass[aps,prl,reprint,superscriptaddress,longbibliography]{revtex4-2}
\usepackage[T1]{fontenc}
\usepackage[utf8]{inputenc}
\usepackage{amsmath,amssymb,bm,graphicx,booktabs}
\usepackage[colorlinks=true,citecolor=blue,linkcolor=blue,urlcolor=blue]{hyperref}
\newcommand{\ii}{\mathrm i}

\begin{document}
\title{Floquet-Volkov spin polarization in  spin-degenerate Antiferromagnets}

\author{Jiyang Fan}
\affiliation{College of Science, China University of Petroleum (East China), Qingdao 266580, China}

\author{Dinghui Wang}
\affiliation{School of Materials Science and Physics, China University of Mining and Technology, Xuzhou 221116, China}
\affiliation{Jiangsu Physical Science Research Center, Nanjing 210093, China}

\author{Tongshuai Zhu}
\email{tongshuaizhu@upc.edu.cn}
\affiliation{College of Science, China University of Petroleum (East China), Qingdao 266580, China}
\begin{abstract}
Periodic driving enables ultrafast control of electronic structures in solids through the formation of photon-dressed Floquet–Bloch states. However, the measured photoemission spectra are influenced not only by these states but also by their interference with Volkov-dressed outgoing photoelectrons. Here, we demonstrate that Floquet–Volkov interference can generate spin-polarized photoemission in a collinear antiferromagnetic Dirac model without spin–orbit coupling (SOC). An in-plane linearly polarized pump preserves  $\mathcal{PT}$ symmetry, leaving the Floquet quasienergies spin degenerate. Under near-resonant driving, photon-dressed conduction and valence bands hybridize, producing spin-dependent relative phases that are converted into unequal spin-resolved photoemission intensities through interference between pathways reaching the same Volkov-dressed final state. We analytically identify this interference mechanism and verify it using nonequilibrium Green-function calculations under periodic driving and finite pump–probe pulses. We further demonstrate that rotating the probe polarization can suppress or reverse the spin polarization without modifying the Floquet bands.  Our findings establish Floquet–Volkov interference as a route to spin-selective photoemission in driven antiferromagnets without Floquet spin splitting or SOC.
\end{abstract}
\maketitle

\emph{Introduction.}---
Periodic driving allows for the ultrafast engineering of electronic structures in solids via photon-dressed Floquet–Bloch states\cite{Sambe1973,Oka2009}, thereby enabling dynamic control over quantum materials, such as  the opening of a surface Dirac gap in Bi$_{2}$Se$_{3}$ under circularly polarized light illumination\cite{Wang2013FloquetTI}, pseudospin-selective band hybridization in black phosphorus\cite{Zhou2023Pseudospin}, helicity-dependent gap control in the topological antiferromagnet MnBi$_{2}$Te$_4$\cite{Bielinski2025Dirac}, the emergence of ultrafast topological states in SnTe\cite{Chassot2026SnTe}, and so on\cite{McIver2020Hall,Zhou2023BelowGap,Bao2024Glide,Wang2026GrapheneGap,Inoue2010,Kitagawa2010,Lindner2011,Kitagawa2011,Rudner2013,Hubener2017,Oka2019,Rudner2020,DeLaTorre2021,zhu2023floquet,liu2018photoinduced,Yan2016Tunable,Bao2024Symmetry,fu2026Floquet,qin2023Light}.

Time- and angle-resolved photoemission spectroscopy (tr-ARPES) has emerged as a powerful technique for probing light-driven electronic structures\cite{Freericks2009,Sentef2015,Aeschlimann2021Scattering,Boschini2024Time,Schmitt2008Transient,Schuler2020}. However, the measured spectra reflect not only the pump-dressed Floquet–Bloch states within the solid but also the Volkov dressing of outgoing photoelectrons\cite{Park2014,Fan2025Selection,Cai2026Occupation,Gadge2026Comparative,Broers2022,MiajaAvila2006,Saathoff2008,Electromagnetic2020Keunecke,Ito2023Subcycle}. When photon-exchange pathways before and after photoemission lead to the same detected energy and in-plane momentum, their transition amplitudes can interfere, giving rise to Floquet–Volkov interference. Such Floquet–Volkov interference has been demonstrated through polarization-dependent sideband intensities in Bi$_2$Se$_3$\cite{Mahmood2016Selective}, controllable momentum asymmetries in black phosphorus~\cite{Bao2025Interference} and graphene\cite{Merboldt2025Graphene,Choi2025Graphene}, and valley-dependent photoemission patterns in WSe$_2$\cite{Fragkos2025Valleytronics}. These findings highlight that photoemission spectral weights encode information about the underlying electronic wave functions beyond that contained in the Floquet quasienergy dispersion, establishing photoemission interference as a promising probe of internal electronic degrees of freedom.

However, the role of Floquet–Volkov interference in spin-dependent photoemission remains largely unexplored. Previous studies have shown that circularly polarized light can induce momentum-dependent Floquet spin splitting in spin-degenerate antiferromagnets by breaking the symmetries protecting spin degeneracy\cite{Zhu2026FloquetOddParity,Huang2026OddParity,Liu2026Dimerized,Li2026FloquetSpin,Zhu2026FloquetSpinGroup,Pan2025Floquet}. Meanwhile, quantum interference between different orbital photoemission channels has been demonstrated to control photoelectron spin polarization in systems with spin–orbit coupling (SOC)\cite{Zhu2014SpinControl,Kuroda2016CoherentSpin,Yaji2017SpinInterference,Park2012SpinPhotoelectrons,Jozwiak2013SpinFlipping,SanchezBarriga2014SpinManipulation,Fanciulli2017SpinDelay,Kuroda2022OpticalSpin}. Nevertheless, whether Floquet–Volkov interference can generate spin-polarized photoemission when the underlying Floquet bands remain spin degenerate and SOC is absent remains an open question.

In this letter, we demonstrate that Floquet–Volkov interference can generate spin-polarized photoemission in a collinear antiferromagnetic Dirac model without SOC. An in-plane linearly polarized pump preserves the $\mathcal{PT}$ symmetry of the system, leaving the Floquet quasienergies spin degenerate and the Floquet-only photoemission signal unpolarized. Under near-resonant driving, photon-dressed conduction and valence bands hybridize. Through the photoemission matrix elements, the probe projects the conduction- and valence-band components of the hybridized states onto a common Volkov-dressed final state channel, allowing the corresponding photoemission pathways to interfere. Owing to the spin-dependent sublattice composition of the antiferromagnetic wave functions, the relative phase between these pathways can differ between the two spin sectors. Consequently, Floquet–Volkov interference converts spin-dependent phase information into unequal spin-resolved photoemission intensities, producing spin-polarized photoelectrons without lifting the spin degeneracy of the Floquet quasienergy bands. We further numerically verify this effect using nonequilibrium Green-function calculations under both periodic driving and finite pump–probe pulses. To facilitate experimental realization, we propose a transverse-magnetic (TM) evanescent field generated by total internal reflection~\cite{Bliokh2012,Bliokh2014,Bliokh2015}, which naturally provides an in-plane field for Floquet driving and an out-of-plane field for Volkov dressing. Our findings reveal a mechanism for generating spin-polarized photoemission without SOC or Floquet spin splitting, providing a  route to probing and controlling spin-dependent quantum interference in driven antiferromagnets.


\begin{figure}[t]
\centering
\includegraphics[width=0.47\textwidth]{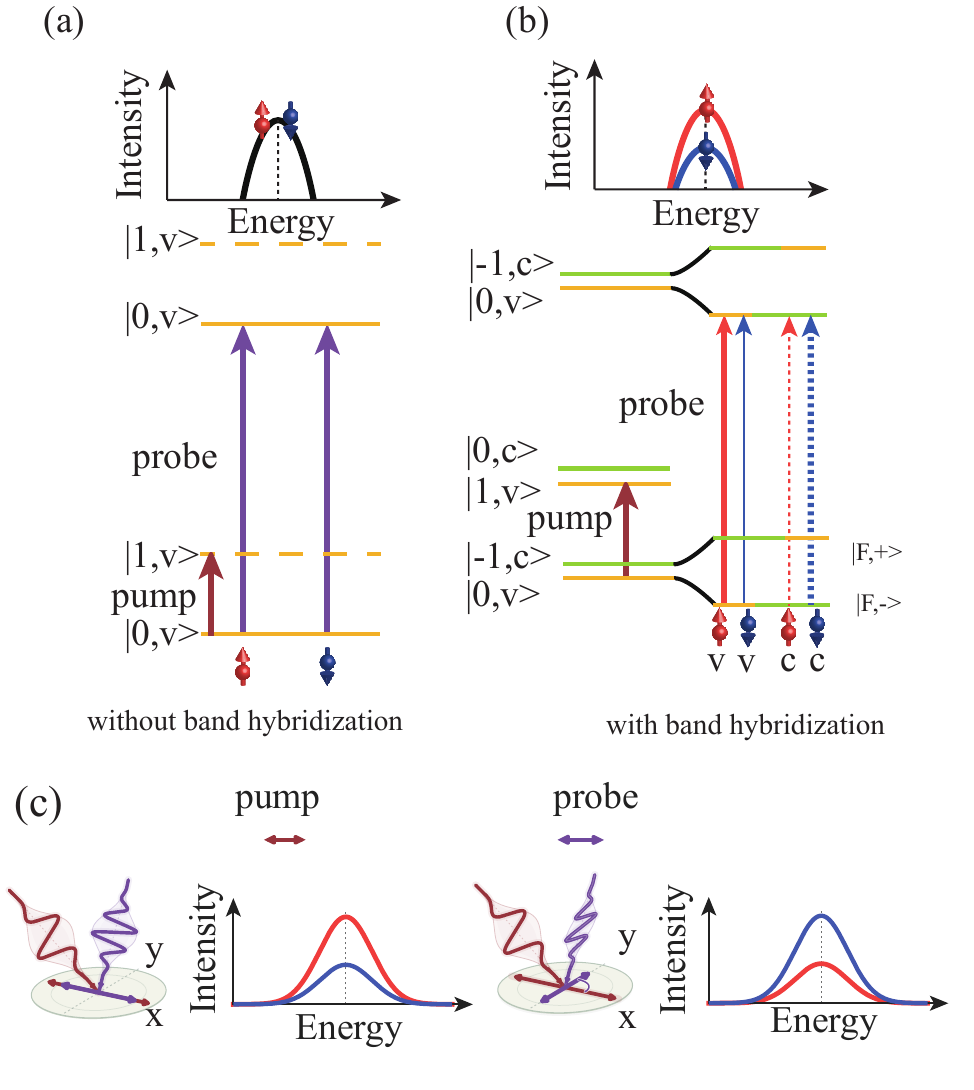}
\caption{Schematic illustration of the effect of Floquet–Volkov interference on photoelectron spin polarization. (a) When the pump generates photon-dressed valence-band replicas without inducing conduction–valence band hybridization, the spin-up and spin-down photoemission intensities remain identical. (b) Near a Floquet avoided crossing, the conduction- and valence-derived Floquet components hybridize. The probe couples these components to a common Volkov-dressed final-state channel through pathways A and B. The spin-dependent relative phase between the two pathways leads to unequal photoemission intensities for the two spin channels. Red and blue arrows denote spin-up and spin-down photoemission, respectively. The initial-state Floquet quasienergies remain spin degenerate in both cases. (c) The photoelectron spin polarization reverses as the relative angle between the pump and probe polarizations is varied.}
\label{fig:mechanism}
\end{figure}

\emph{Mechanism of Floquet–Volkov spin polarization.}---The antiferromagnetic Dirac Hamiltonian is
\begin{equation}
\begin{split}
\mathcal H(\mathbf k)=v(k_x\tau_y+k_y\tau_x)s_0
 +m_z\tau_zs_z,
\end{split}
\label{eq:H}
\end{equation}
\begin{figure*}[t]
\includegraphics[width=\textwidth]{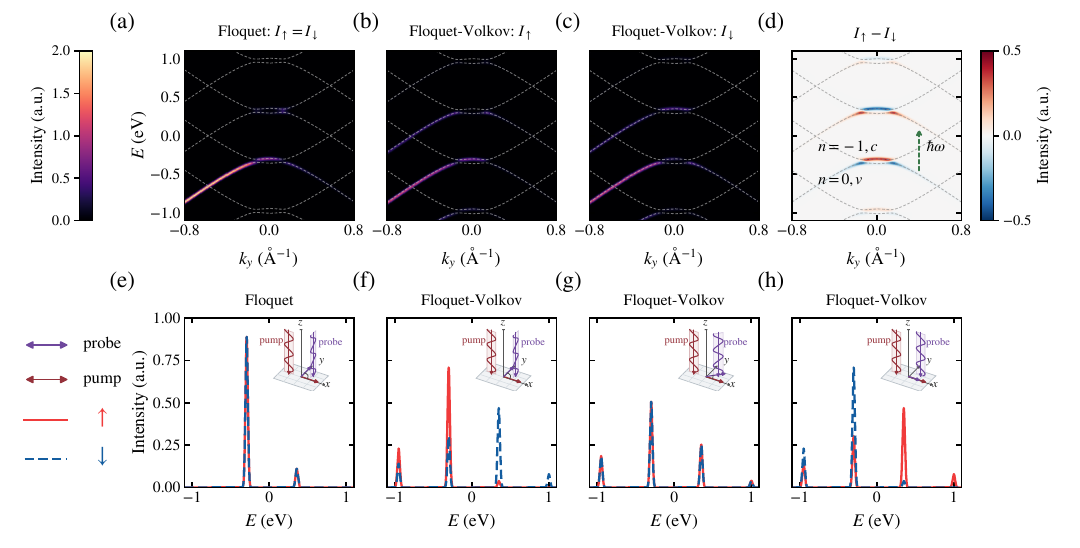}
\caption{spin-resolved photoemission spectra. Upper row: Photoemission spectra calculated under periodic driving at $k_x=0$ and $\alpha=\pi/2$, $\theta=0$. (a) Floquet-only spectra, showing identical photoemission intensities for spin-up and spin-down channels. (b), (c) Spin-up and spin-down photoemission intensities, respectively, including Floquet–Volkov interference. (d) Corresponding spin-resolved intensity difference $I_\uparrow-I_\downarrow$. Gray dashed curves indicate Floquet replicas. Lower row: Photoemission spectra calculated under finite pump–probe pulses at $\mathbf{k}=0$. (e) Spin-resolved spectra including only Floquet dressing. (f)–(h) Floquet–Volkov spectra for probe polarization angles $\alpha=\pi/2$, $\pi/4$, and $0^\circ$, respectively. Red solid and blue dashed curves represent the spin-up and spin-down photoemission intensities, respectively. The pump is polarized along the $x$ direction ($\theta=0$). The parameters are $m_z=0.3$ eV, $\hbar\omega=0.65$ eV, and $a_0=0.04$ \AA$^{-1}$. The pump and probe pulse widths are $T_{\rm p}=250$ and $T_{\rm pr}=30$ fs, respectively, with a pump–probe delay of $t_d=0$ fs.}
\label{fig:spectra}
\end{figure*}
where $\tau_i$ and $s_i$ act on sublattice and spin, respectively,
and $m_z$ is the staggered exchange mass term. The band
energies are $E_{\lambda s}=\lambda\varepsilon_k$, with
$\varepsilon_k=\sqrt{v^2(k_x^2+k_y^2)+m_z^2}$, where $s=\uparrow/\downarrow$, $\lambda=\pm 1$  for conduction and valence band. The polarization directions of the in-plane linearly polarized pump and probe fields are characterized by angles $\theta$ and $\alpha$, respectively, measured from the sample $x$ axis. The vector potential of the linearly polarized pump  is given as $\mathbf A_{\rm pump}(t)=A_0\sin\omega t(\cos\theta\hat{x}+\sin\theta\hat{y})$, where $\theta$ is the polarization directions of the pump light. The driven Hamiltonian for each spin sector is $\mathcal H(\mathbf k,t)=\mathcal H(\mathbf k)+g_F\sin\omega t\,Q_\theta$, where $g_F=va_{0}$, $a_{0}=eA_0/\hbar$, $Q_\theta=(\cos\theta\tau_y+\sin\theta\tau_x)s_{0}$. This pump preserves the  $\mathcal{PT}$ symmetry,
so the Floquet quasienergies remain spin degenerate.  To isolate intraband dressing, we first retain only the band-diagonal pump coupling and neglect
interband transitions. The corresponding Floquet states are
\begin{equation}
\begin{split}
|\Psi^{F(0)}_{\lambda s}(t)\rangle=e^{-\ii\lambda\varepsilon_k t/\hbar}b^\lambda(t)|u_{\lambda s}\rangle,
\end{split}
\label{eq:uncoupledFloquet}
\end{equation}
where $b^\lambda(t)=e^{\ii\lambda\beta\cos\omega t}
 =\sum_n\ii^nJ_n(\lambda\beta)e^{-\ii n\omega t}$ and
$|u_{\lambda s}\rangle$ is the eigenstate of the undriven Hamiltonian. The modulation strength is
$\beta=g_Fq_\theta/(\hbar\omega)$, with
$q_\theta=v(k_x\cos\theta+k_y\sin\theta)/\varepsilon_k$.

The weak probe field, described by $\hat{V}_{\rm pr}(t)=\hat{D} e^{-\ii\omega_{\rm pr}t}$ nduces photoemission into a pump-dressed final state. Neglecting the ponderomotive term for simplicity, the corresponding Volkov state can be written as
\begin{equation}
|\Psi_{\mathbf p}^V(t)\rangle
=|\mathbf p\rangle e^{-\ii E_{\mathbf p}t/\hbar}
 \sum_j a_j e^{-\ii j\omega t}.
\label{eq:Volkov}
\end{equation}
where $\hat{D}=\mathcal{C}(\cos\alpha \tau_{y}+\sin\alpha \tau_x)$ is the probe-induced photoemission transition operator, $|\mathbf p\rangle$ free electron momentum eigenstate, $E_{\mathbf p}=\mathbf{p}^2/(2m_e)$ is the photoelectron energy, where $\mathbf{p}$ is the photoelectron momentum and $m_e$ is the electron mass. $a_{j}$ is the Fourier coefficients of the periodic Volkov phase. For a normal field $\mathbf A(t)=A_z\cos(\omega t)\hat{z}$, $a_j=J_j(\nu_z)$, with
$\nu_z=e k_{fz}A_z/(m_e\omega)$ and $p_z= \hbar k_{fz}$.
The in-plane and out-of-plane pump components are in quadrature. For an initial Floquet state $\psi_s(t)$, the first-order transition amplitude by the Born approximation is\cite{Park2014}
\begin{equation}
\mathcal S_{\mathbf p,s}^{(1)}
=-\frac{\ii}{\hbar}\int dt\,
\langle\Psi^V_{\mathbf p}(t)|\hat{D}_\alpha e^{-\ii\omega_{\rm pr}t}|\psi_s(t)\rangle.
\label{eq:Born}
\end{equation}
Substituting Eq.~(\ref{eq:uncoupledFloquet}) gives the uncoupled $n$th-sideband amplitude
\begin{equation}
\begin{aligned}
\mathcal T^{(0)}_{\lambda s,n}&=M_{\lambda s}\mathcal C_n^\lambda,\\
\end{aligned}\label{eq:uncoupledT}
\end{equation}
where $\mathcal C_n^\lambda=\sum_j a_j^*\ii^{n+j}J_{n+j}(\lambda\beta)$, $M_{\lambda s}=\mathbf M |u_{\lambda s}\rangle$, $\mathbf M=(M_A,M_B)$, $M_a=\langle\mathbf p|\hat D|a\rangle$ ($a=A,B$) denotes the sublattice-resolved photoemission dipole matrix element, with $|a\rangle$ representing the corresponding sublattice basis state. Since the localized orbitals on the two sublattices are related by $\mathcal{PT}$ symmetry in real space, their photoemission matrix elements satisfy $|M_A|=|M_B|$ within the plane-wave final-state approximation. $|\mathcal T^{(0)}_{\lambda\uparrow,n}|^2=|\mathcal T^{(0)}_{\lambda\downarrow,n}|^2$, as in Fig.~\ref{fig:mechanism}(a). In the absence of interband coupling, the photoemission spectrum remains spin unpolarized.

Near the $l$-photon resonance, $2\varepsilon_k=l\hbar\omega$, interband coupling hybridizes the conduction- and valence-derived Floquet replicas. Within the multiphoton rotating-wave approximation, we obtain
\begin{equation}
\begin{aligned}
|\Psi^F_{+s}\rangle&=e^{-\ii\epsilon_+t/\hbar}
 [C_l b^+(t)|u_{+s}\rangle+e^{-\ii\chi_{ls}}S_l e^{\ii l\omega t}b^-(t)|u_{-s}\rangle],\\
|\Psi^F_{-s}\rangle&=e^{-\ii\epsilon_-t/\hbar}
 [e^{-\ii\chi_{ls}}C_l e^{\ii l\omega t}b^-(t)|u_{-s}\rangle-S_l b^+(t)|u_{+s}\rangle],
\end{aligned}\label{eq:hybrid}
\end{equation}
where $\chi_{ls}=\arg V_{ls}$, $\delta_l=2\varepsilon_k-l\hbar\omega$, $V_{ls}$ is the effective $l$-photon interband coupling,  $V_{ls}=-(-i)^l\rho_l\langle u_{+s}|Q_\theta|u_{-s}\rangle$, $\rho_l=g_FlJ_l(2\beta)/(2\beta)$, $\Lambda_l=\sqrt{\delta_l^2+4|V_{ls}|^2}$, $\epsilon_\pm=l\hbar\omega/2\pm\Lambda_l/2$, $C_l^2=(1+\delta_l/\Lambda_l)/2$, and $S_l^2=(1-\delta_l/\Lambda_l)/2$. The resulting photoemission amplitudes are
\begin{equation}
\begin{aligned}
\mathcal T_{+s,n}&=C_lM_{+s}\mathcal C_n^++e^{-\ii\chi_{ls}}S_lM_{-s}\mathcal C_{n+l}^-,\\
\mathcal T_{-s,n}&=-S_lM_{+s}\mathcal C_n^++e^{-\ii\chi_{ls}}C_lM_{-s}\mathcal C_{n+l}^-.
\end{aligned}\label{eq:paths}
\end{equation}
For the hybridized Floquet bands, as illustrated in Fig.~\ref{fig:mechanism}(b), the conduction- and valence-band components of a hybridized Floquet state give rise to two photoemission pathways that reach the same Volkov-dressed final-state channel. The transition amplitudes associated with these pathways are spin dependent. Consequently, their  superposition produces different interference contributions to the spin-resolved photoemission intensities, resulting in a finite spin polarization in the measured photoemission spectrum. For each hybridized branch $\eta$, let $I_{\eta s,n}$ denote the photoemission intensity of the $n$th sideband in spin channel $s$. The corresponding spin-resolved photoemission intensity difference is defined as $\Delta I_{\eta,n}=I_{\eta\uparrow,n}-I_{\eta\downarrow,n}$. Using Eq.~(\ref{eq:paths}), we obtain
\begin{equation}
\begin{aligned}
\Delta I_{\eta,n}&\propto
 |\mathcal T_{\eta\uparrow,n}|^2-|\mathcal T_{\eta\downarrow,n}|^2\\
&=-\frac{2\eta\rho_lm_z|\mathcal C|^2}{\Lambda_l\varepsilon_k}\cos(2\alpha-\theta)
 \operatorname{Im}\!\left[\ii^l
 (\mathcal C_n^+)^*\mathcal C_{n+l}^-\right],\\
\end{aligned}\label{eq:intensity}
\end{equation}
with a finite relative phase between the in-plane and out-of-plane components of the pump light, both the magnitude and sign of the photoelectron spin polarization can be controlled by tuning the polarization angles of the pump and probe beams. A detailed derivation of the spin-dependent interference term and the resulting photoemission intensity difference is provided in the Supplemental Material~\cite{SM}.

\emph{Spin-resolved  tr-ARPES spectra.}---We calculate the spin-resolved tr-ARPES signal from the lesser Green function $G_s^<(\mathbf k, t,t')$~\cite{Freericks2009,Sentef2015,Gadge2026Comparative}:
\begin{equation}
\begin{aligned}
I_s(E)\propto\operatorname{Im}\int dt\,dt'\,\mathcal K_E(t,t')
\mathbf MG_s^<(\mathbf k, t',t)\mathbf M^\dagger,
\end{aligned}\label{eq:NEGF}
\end{equation}
where $\mathcal K_E(t,t')=g(t)g(t')e^{-\ii \varphi(\mathbf k,t,t')}$, $\varphi(\mathbf k, t,t')=\frac{1}{\hbar}\int_{t'}^{t}d \overline{t}[\varepsilon_f(\overline{t})-\hbar\omega_{\rm pr}]$, $\varepsilon_f({t})$ is the light-dressed final state energy and $g(t)=e^{-(t-t_d)^2/(2T_{\rm pr}^2)}/(\sqrt{2\pi}T_{\rm pr})$ with $t_{d}$ is the delay time between the pump and probe peaks, $T_{\rm pr}$ is the probe pulse width.

Figure~\ref{fig:spectra}(a) presents the photoemission spectra obtained by considering only Floquet dressing, for which the spin-up and spin-down spectral intensities are identical. Figures~\ref{fig:spectra}(b)–(d) show the spin-up and spin-down photoemission spectra, as well as their intensity difference $I_\uparrow-I_\downarrow$, respectively, when Floquet–Volkov interference is included. The pump photon energy is set to \(\hbar\omega=0.65\) eV. Pronounced spin polarization emerges near the avoided crossings due to strong hybridization between the Floquet bands. To further elucidate the roles of photoemission matrix elements and Floquet–Volkov interference, we calculate the energy-dependent spin-resolved tr-ARPES spectra at $\mathbf{k}=0$. With Floquet dressing alone[\ref{fig:spectra}(e)], the photoemission spectra remain spin unpolarized even when matrix-element effects are included. In Figs.~\ref{fig:spectra}(f)–(h), we include both Floquet–Volkov interference and photoemission matrix-element effects, considering relative polarization angles of $\pi/2$, $\pi/4$, and $0$ between the pump and probe fields, respectively. The results demonstrate that varying the relative polarization angle can effectively control the spin polarization, including its complete suppression and reversal, in agreement with the analytical prediction of Eq.~(\ref{eq:intensity}).

\begin{figure}[t]
\centering
\includegraphics[width=\columnwidth]{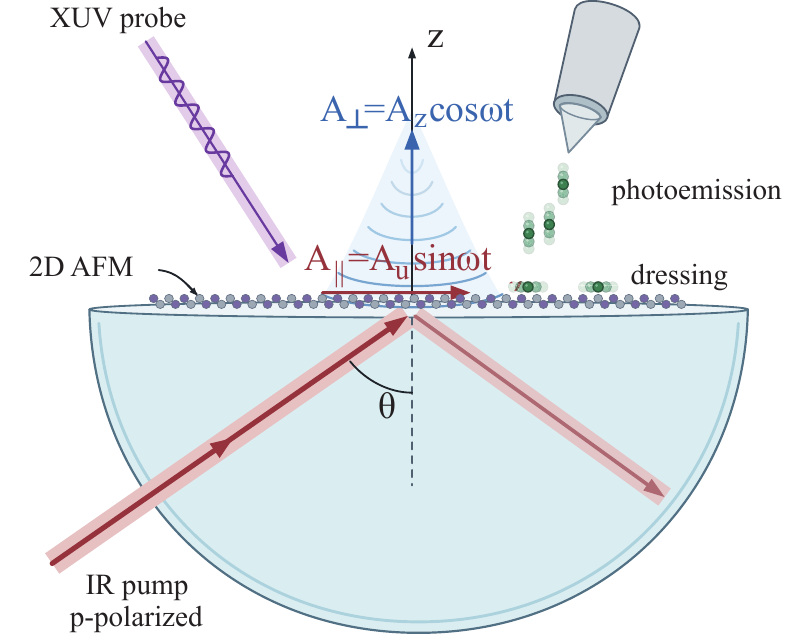}
\caption{Schematic of the experimental setup. A linearly p-polarized infrared pump beam undergoes total internal reflection at a prism–vacuum interface supporting a two-dimensional antiferromagnet. The resulting transverse-magnetic evanescent field possesses tangential and normal components in phase quadrature. An independent extreme-ultraviolet (XUV) probe pulse excites and releases photoelectrons into this field for spin-resolved detection.}
\label{fig:optics}
\end{figure}

\emph{Experimental setup.}---The optical configuration illustrated in Fig.~\ref{fig:optics} realizes the proposed field geometry through total internal reflection of a $p$-polarized infrared pump.  The resulting TM evanescent resulting vector potential can be expressed as $\mathbf A_{\mathrm{ev}}(z,t)
=A_0e^{-z/\ell_A}
\left[
\hat{\mathbf x}\sin(\omega t)
-\xi\hat{\mathbf z}\cos(\omega t)
\right]$\cite{Bliokh2012,Bliokh2014,Bliokh2015},
where $\ell_A={c}/({\omega\sqrt{n_1^2\sin^2\theta_i-1}})$ is the evanescent decay constant, $\xi={n_1\sin\theta_i}/{\sqrt{n_1^2\sin^2\theta_i-1}}$ is the amplitude ratio between the normal and tangential components,  $n_1$ and $\theta_i$ denoting the prism refractive index and incidence angle, respectively. The evanescent field propagates along the interface and decays exponentially along the surface normal. Within the two-dimensional material, the tangential component of the optical field drives the bound electrons into Floquet states, whereas the normal component induces Volkov dressing of the photoelectrons emitted by the probe pulse. The prism refractive index and incidence angle jointly determine the evanescent decay length and the surface amplitude ratio between the tangential and normal field components. Moreover, the spatial decay of the optical field modifies the dynamic phase accumulated by photoelectrons escaping from the surface. For $n_1=1.5$, $\theta_i=50^\circ$, and $\hbar\omega=0.65$~eV, these expressions give
$\ell_A\simeq536$~nm and $\xi\simeq2.03$. We suppose only the normal component
contributes to the  outgoing photoelectron and using the straight trajectory $z=v_{fz}\tau$, where
$v_{fz}=\hbar k_{fz}/m_e$, gives $J_{1,\ell}(t)
=
\nu_z\omega\int_0^\infty d\tau\,
e^{-\gamma\tau}\cos[\omega(t+\tau)]
=
-\nu_{z,\ell}\sin(\omega t-\phi_{\rm esc})$, where $\gamma={v_{fz}}/{\ell_A}$, $\nu_z=e v_{fz}A_z/(\hbar\omega)$, $\nu_{z,\ell}
={\nu_z}/{\sqrt{1+\eta_{\rm esc}^2}}$, $\phi_{\rm esc}=\arctan\frac{v_{fz}}{\omega\ell_A}$. Spatial decay therefore reduces the accumulated phase-modulation amplitude and introduces a phase shift relative to the fixed surface field. For a photoelectron kinetic energy of $20$ eV, $v_{fz}\simeq2.652$ nm/fs and $\eta_{\rm esc}\simeq5.01\times10^{-3}$. The corresponding  phase shift is $0.287^\circ$, indicating that the effect of spatial decay on the outgoing phase modulation remains small for the considered geometry.

\emph{Conclusion.}---In summary, we have revealed a mechanism for spin-polarized photoemission induced by Floquet-Volkov quantum interference in collinear antiferromagnets. In the absence of SOC, an in-plane linearly polarized pump field preserves the  $\mathcal{PT}$ symmetry of the system, keeping the Floquet quasienergy bands spin-degenerate such that photoemission signals arising solely from Floquet dressing exhibit no spin polarization. Under near-resonant driving, Floquet replicas of the conduction and valence bands hybridize, and their distinct band components couple to a common Volkov-dressed final state via two photoemission pathways. Owing to the spin-dependent sublattice composition of the antiferromagnetic wave functions, the relative phase between these two pathways differs across opposite spin channels, thereby generating spin-polarized photoemission through quantum interference. By tuning the relative polarization angle between the pump and probe pulses, one can effectively control the magnitude of the spin polarization and reverse its sign without lifting the spin degeneracy of the Floquet bands. Experimentally, we propose utilizing a TM evanescent field generated via total internal reflection to simultaneously realize in-plane Floquet driving and out-of-plane Volkov dressing, which can be directly observed via spin-resolved tr-ARPES. Van der Waals magnets MnPX$_3$ ($X=\text{S}, \text{Se}$)\cite{Yang2020MnPX3,Chu2020MnPS3,Gao2023MnPS3}, the Dirac semimetal CuMnAs\cite{Tang2016CuMnAs,Wadley2016CuMnAs,Smejkal2017CuMnAs}, and even-layer MnBi$_2$Te$_4$\cite{Zhang2019Axion,Otrokov2019MnBi2Te4,Li2019MnBi2Te4}  serve as viable material platforms to validate our theoretical predictions. Our findings demonstrate that spin polarization in photoemission does not necessarily imply spin splitting in the initial-state band structure, but can instead originate directly from coherent quantum interference during the photoemission process. This work opens a new avenue for achieving ultrafast spin-selective photoemission and probing spin-dependent quantum coherence in SOC-free antiferromagnetic systems.

This work is supported by the National Natural Science Foundation of China (Grants No. 12504197 and 12604376), Shandong Provincial Natural Science Foundation (Grant No. ZR2024QA095), the Youth Innovation Team Plan Project for the Higher Education Institution of Shandong Province (Grant No. 2024KJN021), the Fundamental Research Funds for the Central Universities (Grant No. 2024QN11046), and Natural Science Foundation of Jiangsu Province (Grant No. BK20233001). We acknowledge the assistance of Codex and ChatGPT 6.0 with manalytical derivations, figure preparation, and language editing during the preparation of this manuscript.

\bibliography{references}
\end{document}